\documentclass{article}
\usepackage[preprint]{spconf}
\usepackage{amsmath, graphicx}
\usepackage{amssymb}
\usepackage{xcolor}
\usepackage{multirow}
\usepackage[export]{adjustbox}
\usepackage{makecell}
\usepackage{tablefootnote}
\usepackage{arydshln}
\usepackage[hidelinks]{hyperref}
\hypersetup{
    colorlinks=true,
    linkcolor=black,
    citecolor=black,
}

\newcommand\ie{\textit{i.e.}}
\newcommand\eg{\textit{e.g.}}

\def\qzi{ q_{Z_i | X, Z_{<i}} }
\def\pzi{ p_{Z_i | Z_{<i}} }
\def\cbase{ C }

\def\ourfunction{smoothing function}

\newcommand\blfootnote[1]{%
  \begingroup
  \renewcommand\thefootnote{}\footnote{#1}%
  \addtocounter{footnote}{-1}%
  \endgroup
}

\title{An Improved Upper Bound on the Rate-Distortion Function of Images}

\name{Zhihao Duan\textsuperscript{*}, Jack Ma\textsuperscript{*}, Jiangpeng He, and Fengqing Zhu
}
\address{
\textsuperscript{*}Equal contribution, in alphabetical order \\
Elmore Family School of Electrical and Computer Engineering, \\
Purdue University, West Lafayette, Indiana, U.S.A.
}

\begin{document}
\maketitle

\begin{abstract}

Recent work has shown that Variational Autoencoders (VAEs) can be used to upper-bound the information rate-distortion (R-D) function of images, \ie, the fundamental limit of lossy image compression.
In this paper, we report an improved upper bound on the R-D function of images implemented by (1) introducing a new VAE model architecture, (2) applying variable-rate compression techniques, and (3) proposing a novel \ourfunction{} to stabilize training.
We demonstrate that at least 30\% BD-rate reduction w.r.t. the intra prediction mode in VVC codec is achievable, suggesting that there is still great potential for improving lossy image compression.
Code is made \href{https://github.com/duanzhiihao/lossy-vae}{publicly available}.
\end{abstract}

\blfootnote{
\copyright 2023 IEEE. Personal use of this material is permitted. Permission from IEEE must be obtained for all other uses, in any current or future media, including reprinting/republishing this material for advertising or promotional purposes, creating new collective works, for resale or redistribution to servers or lists, or reuse of any copyrighted component of this work in other works.
}

\section{Introduction}

\begin{figure}[t]
    \centering
    \includegraphics[width=0.8\linewidth]{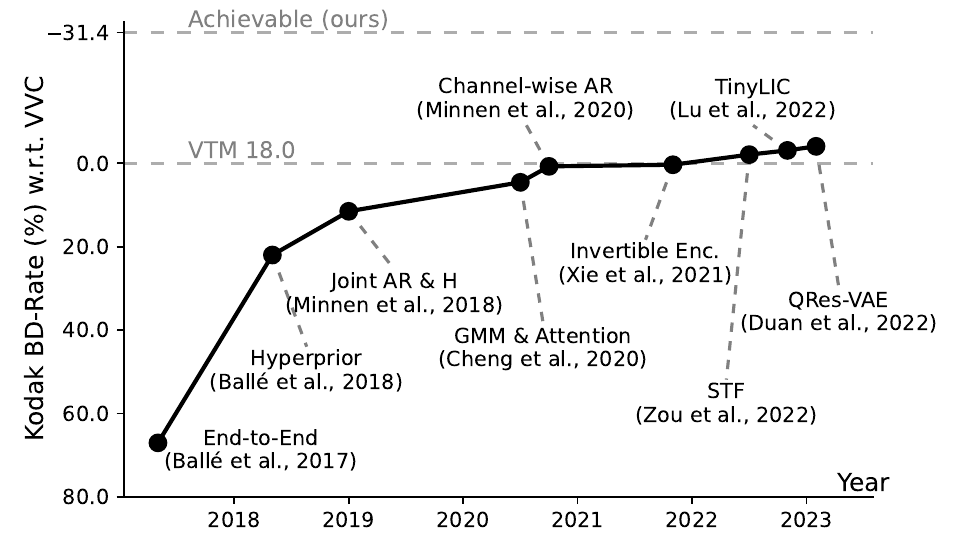}
    \vspace{-10pt}
    \caption{
    Evolution of learned image compression methods.
    In this paper, we show that more than 30\% BD-rate reduction w.r.t. VVC intra codec~\cite{pfaff2021vvc_intra} (VTM version 18.0) is achievable, suggesting that recent methods are still far from optimum.}
    \vspace{-10pt}
    \label{fig:rd_intro_trend}
\end{figure}

Lossy image compression is one of the most fundamental problems in image processing.
Recent years have viewed a rapid development of lossy image compression systems powered by deep learning techniques, but this trend of improvement has started to stagnate, as shown in Fig.~\ref{fig:rd_intro_trend}.
This observation leads to an interesting question: \textit{what is the gap between current image codecs and the theoretical limit of lossy compression?}
If current algorithms are already near the limit, then further research in lossy image compression could only bring tiny outcomes and thus is not worthwhile.
However, if there is potential for a large improvement, it is still worthwhile to continue searching for better compression techniques.

To address this question, we must measure the fundamental limit of lossy image compression, \ie, the information rate-distortion (R-D) function~\cite{cover2005information}. The information R-D function of a data source $X\sim p_\text{data}$ is given by
\begin{equation}
\label{eq:rd_intro_rd_function}
    R(D) = \min_{p_{\hat{X} | X}} I(X, \hat{X}) \quad
    \text{s.t.} \ \ \mathbb{E}\left[ d(X, \hat{X}) \right] \le D ,
\end{equation}
where $d$ is a distortion metric, the expectation is w.r.t. $X\sim p_\text{data}$ and $\hat{X} \sim p_{\hat{X} | X}$, and the minimization is calculated over all possible distributions for $p_{\hat{X} | X}$.
The function $R(D)$ describes the minimum rate (\ie, the average number of bits for each sample) required to compress an i.i.d. sequence of $X$ under a distortion threshold $D$.
Any R-D pair that lies above $R(D)$ is guaranteed to be achievable; in other words, there exists a codec that can reach this R-D pair.

Despite its importance, computing $R(D)$ is notoriously difficult~\cite{cover2005information}, especially for complex sources such as natural images where $p_\text{data}$ does not have a closed-form expression.
Nevertheless, recent research~\cite{alemi2018fixing_elbo, yang2022sandwich} proved that one could compute a tight upper bound for $R(D)$ using variational autoencoders~\cite{kingma14vae} (VAEs), a family of probabilistic models closely related to information theory.
VAEs provide a set of rate-distortion pairs lying above $R(D)$, and by comparing existing codecs against these pairs, we can assess at least how far the existing codecs are from $R(D)$.
Using this method, Yang and Mandt~\cite{yang2022sandwich} proved that it is possible to improve existing codecs by at least one dB in PSNR at various rates.


\textbf{Relation to prior work and our contributions:}
In this work, we begin by reviewing the theoretical analysis of VAEs and R-D theory adapted from \cite{yang2022sandwich}. We then identify a simple yet useful way to improve the R-D bound: scaling up the VAE model in both the number of channels and the number of layers (\textbf{Contribution 1}).
Since more VAE layers introduce larger gradient noise in training (because each layer involves stochastic sampling), we propose a new \ourfunction{} for the posterior and prior distributions in our model to stabilize training (\textbf{Contribution 2}).
We also adopt variable-rate compression methods to achieve a continuous R-D upper bound function (\textbf{Contribution 3}).
Our results show that lossy image compression still has a large room for improvement. Specifically, more than 30\% BD-rate reduction w.r.t. the VVC intra mode~\cite{pfaff2021vvc_intra} is achievable (\textbf{Contribution 4}).



\section{Related Works}

\textbf{Lossy Image Compression}: Researchers have made enormous progress on lossy image compression over the past few decades. For instance, the discrete cosine transform~\cite{admed1974dct} enabled the development of the JPEG~\cite{wallace1992jpeg}, the currently most popular image codec. In addition to handcrafted codecs, researchers are seeking to apply the potential machine learning brings to the field of image compression.
Recent work has focused primarily on Variational Autoencoders (VAEs) due to their close relationship with data compression and information theory.
Recent learning-based methods~\cite{zou2022stf, lu2022tinylic, duan2023qres, duan2023qarv} all report a better compression performance than the VCC intra codec~\cite{pfaff2021vvc_intra}, perhaps the current best-performing handcrafted image codec.\\\\
\textbf{Estimating the Rate-Distortion Function}: The Blahut-Arimoto algorithm~\cite{blahut1972rd} provides a method for computing the R-D function; however, the algorithm only considers discrete sources with known distributions. 
Harrison and Kontoyiannis~\cite{harrison2008rd} remove these restrictions by considering any general source with an unknown distribution, but their work attacks the image compression problem from an exclusively theoretical standpoint. As for empirically approximating the R-D function for an image source, VAEs provide a promising platform due to their close relationship with rate-distortion theory~\cite{alemi2018fixing_elbo}.
Yang and Mandt~\cite{yang2022sandwich} upper-bound the R-D function for image sources by training VAEs (we overview this method in Sec.~\ref{sec:rd_method_theoretical}).
In doing so, they show that there still exists a sizeable room for improvement over current compression algorithms (one dB in PSNR at various rates).

\section{Method}
We provide an overview of our model in Sec.~\ref{sec:rd_method_arch}, discuss how to obtain the upper bound for the R-D function in Sec.~\ref{sec:rd_method_theoretical}, and present implementation details in Sec.~\ref{sec:rd_method_implementation}.

\subsection{Architecture Overview}
\label{sec:rd_method_arch}
\begin{figure}[t]
    \centering
    \includegraphics[width=0.96\linewidth]{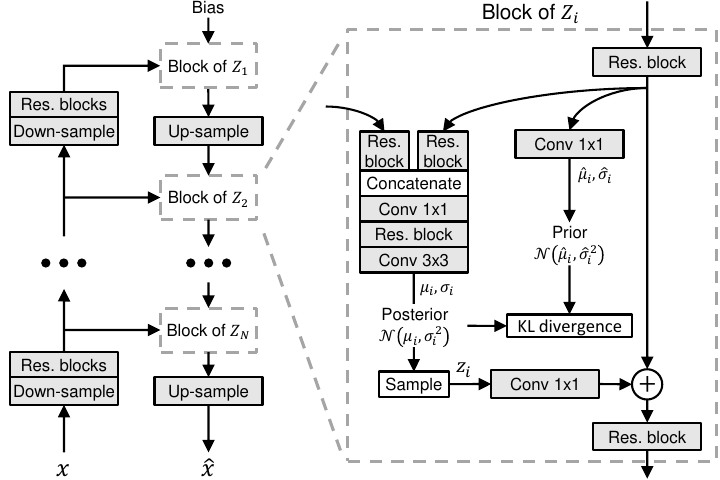}
    \vspace{-8pt}
    \caption{Overview of our model architecture.}
    \label{fig:rd_method_overview}
\end{figure}

Our model architecture is overviewed in Fig.~\ref{fig:rd_method_overview}.
As with many modern VAE models~\cite{vahdat2020nvae, child2021vdvae}, our model employs a ResNet VAE~\cite{kingma2016iafvae} architecture containing a set $Z \triangleq \{ Z_1, Z_2, ..., Z_N \}$ of latent variables ordered by increasing dimensionality.
Latent variables are connected to each other in an autoregressive manner in the sense that each $Z_i$ is conditionally dependent on $Z_{<i} \triangleq \{ Z_1, Z_2, ..., Z_{i-1} \}$ in both inference and sampling.
We use $\qzi$ and $\pzi$ to denote the posterior and prior distributions respectively for a latent variable $Z_i\in Z$.

\subsection{Upper-Bounding the Rate-Distortion Function}
\label{sec:rd_method_theoretical}
Recall that the R-D function $R(D)$, as referenced in Eq. (\ref{eq:rd_intro_rd_function}), is derived by finding
the minimum $I(X, \hat{X})$ given the constraint $\mathbb{E}[ d(X, \hat{X}) ] \le D$.
Although the problem is difficult to solve directly, it has been shown that (\eg, by Sullivan and Wiegand~\cite{sullivan1998rdo_video}) one can apply a Lagrangian relaxation to it by replacing the inequality constraint with a penalty term $\lambda \cdot \mathbb{E} [ d(X, \hat{X}) ]$ in the objective, where $\lambda$ is the Lagrange multiplier.
In doing so, we obtain an unconstrained objective:
\begin{equation}
\label{eq:rd_method_rdo}
    \min_{p_{\hat{X} | X}} \
    I(X; \hat{X}) + \lambda \cdot \mathbb{E} \left[ d(X, \hat{X}) \right],
\end{equation}
and each solution to Eq.~\eqref{eq:rd_method_rdo} for a given $\lambda > 0$ corresponds to a solution to $R(D)$ for a particular value of $D$.
To upper-bound Eq.~\eqref{eq:rd_method_rdo} using VAEs, we use the fact that~\cite{yang2022sandwich}
\begin{equation}
\label{eq:rd_method_upper_bound}
\begin{aligned}
    I(X;\hat{X})
    \le & \, I(X;Z)
    \\
    = & \, \mathbb{E} \left[ D_\text{KL}( q_{Z|X} \parallel q_Z ) \right]
    \\
    \le & \, \mathbb{E} \left[ D_\text{KL}( q_{Z|X} \parallel q_Z ) \right] + D_\text{KL}( q_Z \parallel p_Z )
    \\
    = & \, \mathbb{E} \left[ D_\text{KL}( q_{Z|X} \parallel p_Z ) \right]
    \\
    = & \, \mathbb{E} \left[ \sum_{i=1}^N D_\text{KL} ( \qzi \parallel \pzi ) \right],
\end{aligned}
\end{equation}
where $q_Z$ is the true marginal distribution of $Z$, \ie, the set of all latent variables.
In Eq.~\eqref{eq:rd_method_upper_bound}, 
the first inequality is an application of the data processing inequality, and the second inequality follows because KL divergence is non-negative.
 Adding the penalty term to both sides of Eq.~\eqref{eq:rd_method_upper_bound}, we get:
\begin{equation}
\label{eq:rd_method_loss_bound}
\begin{aligned}
    & \, I(X; \hat{X}) + \lambda \cdot \mathbb{E} \left[ d(X, \hat{X}) \right]
    \\
    \le & \, \mathbb{E} \left[
    \sum_{i=1}^N D_\text{KL} ( \qzi \parallel \pzi )
    +
    \lambda \cdot d(X, \hat{X})
    \right].
\end{aligned}
\end{equation}
We are now ready to set our training loss function as:
\begin{equation}
\label{eq:rd_method_loss}
    \mathcal{L}
    = \mathbb{E} \left[ \sum_{i=1}^N D_\text{KL} ( \qzi \parallel \pzi ) + \lambda \cdot d(X, \hat{X}) \right].
\end{equation}
By minimizing $\mathcal{L}$, an upper bound on the relaxed objective of the R-D function (Eq.~\eqref{eq:rd_method_rdo}), we impose the model to find the best set of parameters such that $\mathcal{L}$ approaches from above the objective of Eq.~\eqref{eq:rd_method_rdo} as closely as possible.
Once training is completed for a particular $\lambda$, we define $D$ and $U(D)$ as
\begin{equation}
\label{eq:rd_method_d_and_u}
\begin{aligned}
    D &\triangleq \mathbb{E} \left[ d(X, \hat{X}) \right]
    \\
    U(D) &\triangleq \mathbb{E} \left[ \sum_{i=1}^N D_\text{KL} ( \qzi \parallel \pzi ) \right].
\end{aligned}
\end{equation}
Then, we have $U(D) \ge I(X; \hat{X}) \ge R(D)$, where the second inequality follows because $R(D)$ is a global minimum of $I(X; \hat{X})$ w.r.t. all possible $p_{\hat{X}|X}$.
Thus, for any $D$, the point $(U(D), D)$ on the R-D plane lies above the point $(R(D), D)$, and by varying $\lambda$ continuously, we can traverse the function $U(D)$, which is an upper bound on the true R-D function.


\begin{table}[t]
\centering
\small
\begin{tabular}{l|ccccc}
\hline
Feature resolution     & 16x16     & 8x8       & 4x4       & 2x2       & 1x1 \\ \hline
\# of channels         & $2\cbase$ & $4\cbase$ & $5\cbase$ & $6\cbase$ & $6\cbase$ \\
\# of latent variables & 5         & 4         & 3         & 2         & 1         \\ \hline
\end{tabular}
\vspace{-0.2cm}
\caption{
Model configuration. Feature resolutions are w.r.t. a $64\times 64$ input image.
The base number of channels, $\cbase$, is 128.}
\label{table:rd_method_config}
\end{table}

\subsection{Model Details}
\label{sec:rd_method_implementation}



\textbf{Number of channels and latent variables}:
As we discussed in Sec.~\ref{sec:rd_method_theoretical}, the minimization in the original R-D function (Eq.~\eqref{eq:rd_intro_rd_function}) is w.r.t. all possible distributions of $p_{\hat{X} | X}$, while in practice the minimization is w.r.t. neural network parameters.
This results in a gap between the best $p_{\hat{X} | X}$ formed by our model and the optimal solution to the optimization problem in the R-D function.
We notice that this gap can be narrowed by scaling up the model's capacity, \ie, increasing its depth (number of latent variables) and width (number of channels), which we empirically show in Sec.~\ref{sec:rd_exp_additional}.
Our final choice of these hyperparameters is depicted in Table~\ref{table:rd_method_config}.

\textbf{Continuous R-D bound by variable-rate training}: 
By using the variable-rate compression method in~\cite{duan2023qarv}, we are able to train a single model that obtains a continuous $U(D)$ function.
As a brief overview, at each training step, a random $\lambda$ is sampled and used in the loss function (Eq.~\eqref{eq:rd_method_loss}).
We sample $\lambda \in [4, 2048]$ uniformly in the log space.
In other words, we first sample an intermediate variable $\lambda' \sim U(\log 2, \log 2048)$, and  assign $\lambda \leftarrow \exp{\lambda'}$.
The model is conditional on $\lambda$ by using the sinusoidal embedding~\cite{vaswani2017attention} and the adaptive Layer Normalization~\cite{ba2016layernorm, duan2023qarv}.
Once training is completed, the model is able to traverse the rate-distortion curve by continuously varying the $\lambda$ input to the model.

\textbf{Training stabilization}:
Following previous works~\cite{vahdat2020nvae, child2021vdvae}, we set all posterior and prior distributions to be (conditional) Gaussian, as shown in Fig.~\ref{fig:rd_method_overview}. This way, the KL divergence term between the posterior and prior for a latent variable $Z_i$ has a closed-form solution:
\begin{equation}
\label{eq:rd_method_gaussian_kl}
    D_\text{KL}( \qzi \parallel \pzi)
    = -\frac{1}{2} + \log \frac{\hat{\sigma}_i}{\sigma_i} +
    \frac{ \sigma_i^2 + (\mu_i - \hat{\mu}_i)^2 }{ 2 \hat{\sigma}_i^2 },
\end{equation}
where $\mu_i, \hat{\mu}_i, \sigma_i, \hat{\sigma}_i$ are the neural network outputs shown in Fig.~\ref{fig:rd_method_overview}.
However, Hierarchical VAEs often produce unbounded gradients resulting from the KL term during training~\cite{child2021vdvae}, giving rise to unstable training. We observe in our experiments that large gradients often come from the $(\mu_i - \hat{\mu}_i)^2$ term, which produces unbounded gradients w.r.t. neural network parameters when $\mu_i$ deviates from $\hat{\mu}_i$.
To combat this issue, we compute $\mu_i$ from network outputs as:
\begin{equation}
    \mu_i = \text{sign}(a_i) \cdot |a_i|^{1 - 0.5 \cdot \tanh(|a_i|)},
\end{equation}
where $a_i$ denotes the output from a convolutional layer. The same smoothing function is also used for $\hat{\mu}_i$.
When $a_i \rightarrow \infty$, we have $\mu_i \rightarrow \sqrt{a_i}$, and the derivative of $(\mu_i - \hat{\mu}_i)^2$ w.r.t. $a_i$ converges to $1$.
The same conclusion holds for $a_i \rightarrow -\infty$.
We assume that by reducing the gradient variance during training, this function improves training convergence and model performance, which we empirically verify in Sec.~\ref{sec:rd_exp_additional}.

\begin{figure*}[ht]
\centering
\begin{minipage}[b]{.32\linewidth}
    \centering
    \centerline{\includegraphics[width=\linewidth]{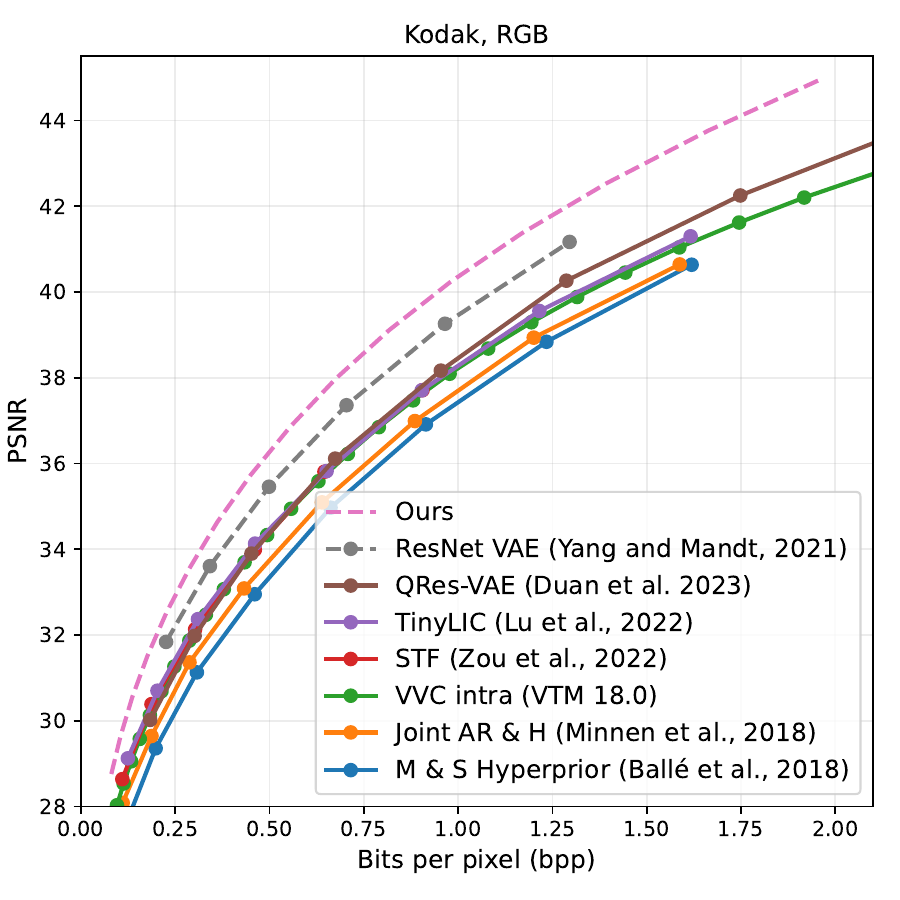}}
\end{minipage}
\begin{minipage}[b]{.32\linewidth}
    \centering
    \centerline{\includegraphics[width=\linewidth]{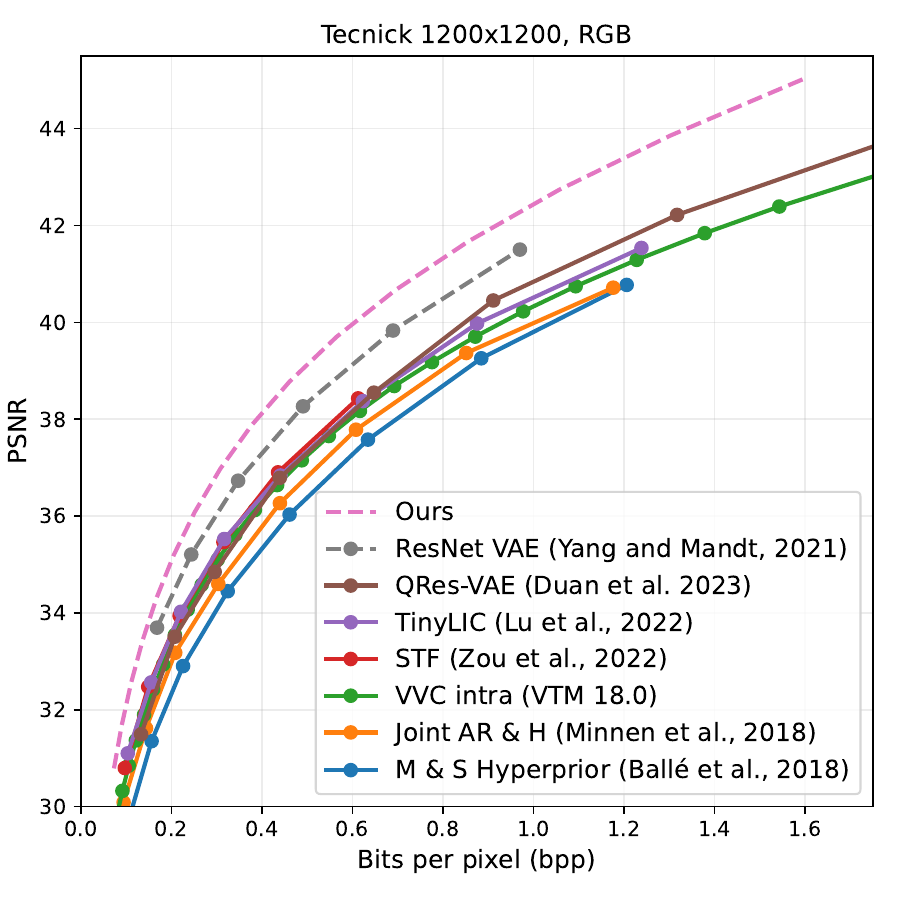}}
\end{minipage}
\begin{minipage}[b]{.32\linewidth}
    \centering
    \centerline{\includegraphics[width=\linewidth]{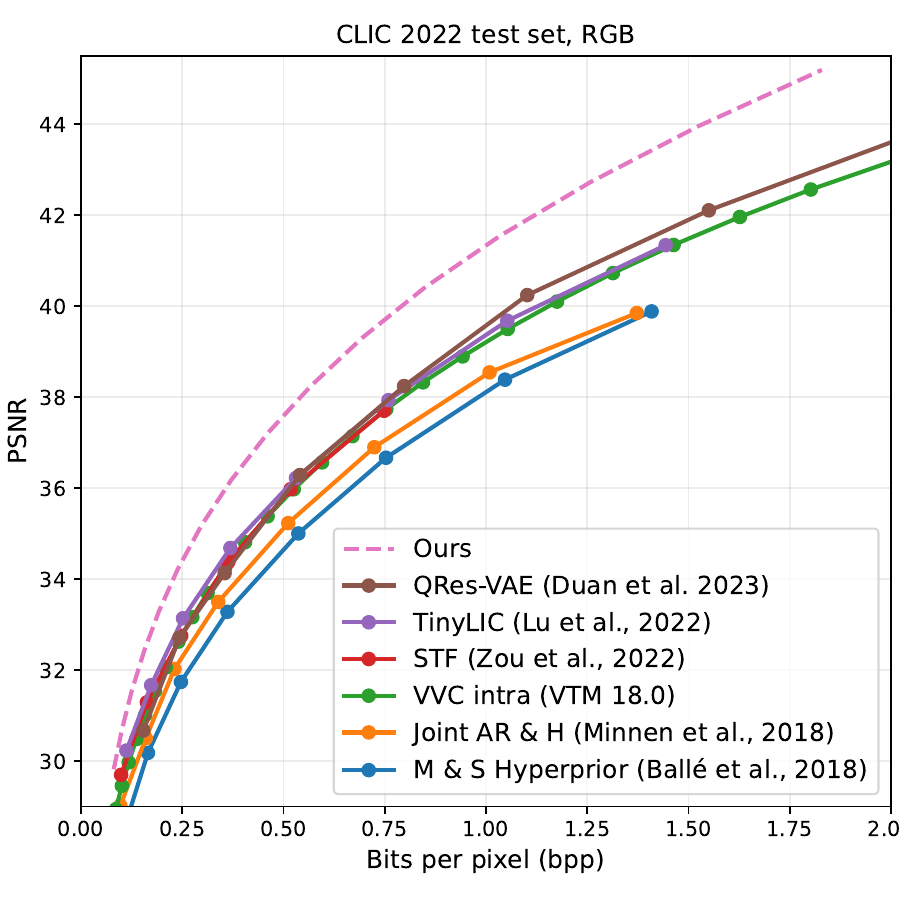}}
\end{minipage}
\vspace{-0.5cm}
\caption{
Comparison between our theoretical bound (a single model achieving a continuous PSNR-bpp curve) and previous methods (separate models for different PSNR-bpp pairs).
Dashed lines denote theoretically achievable performance, and solid lines denote practical image codecs.
Our results show that lossy image compression still has a large room for improvement.}
\label{fig:rd_exp_main}
\vspace{-0.2cm}
\end{figure*}

\section{Experiments}

\subsection{Datasets, Metrics, and Training Details}
In all our experiments, we train our models on the COCO dataset~\cite{lin2014coco} for 200k iterations with batch size 32, Adam optimizer~\cite{Kingma15adam}, and learning rate $2 \times 10^{-4}$.
Mean squared error (MSE) is used to measure distortion $d(X, \hat{X})$.
Other training settings are the same as in~\cite{duan2023qarv} and can be found in our code.

We use three common test sets for evaluation: Kodak~\cite{kodak} (24 images), CLIC\footnote{http://compression.cc} 2022 test partition (30 images), and Tecnick TESTIMAGES~\cite{asuni2014tecnick} (100 images).
We use the pre-trained model to compute $D$ and $U(D)$ as described in Eq.~\eqref{eq:rd_method_d_and_u}, which together form the R-D upper bound.
As per convention, we convert $U(D)$ into bits per pixel (bpp)\footnote{We use natural logarithm for KL, so $U(D)$ has a unit of nats. We convert $U(D)$ into bpp by multiplying $\log_2 e$ and then averaging over image pixels.}, and convert MSE into PSNR\footnote{$\text{PSNR} = -10 \cdot \log_{10} \text{MSE}$}.
BD-rate~\cite{bjontegaard2001bdrate} is used to measure the average rate difference between PSNR-bpp curves.

\subsection{Main Results}
We report the achievable R-D bound obtained by our model in the form of PSNR-bpp curves\footnote{Because PSNR is inversely related to distortion, our PSNR-bpp curve serves as a lower bound for the PSNR-bpp representation of $R(D)$.} in Fig.~\ref{fig:rd_exp_main} and BD-rate in Table~\ref{table:rd_exp_bdrate}.
In addition to our method, we also show results for the previous R-D upper bound~\cite{yang2022sandwich}, learning-based image codecs~\cite{lu2022tinylic, zou2022stf, minnen2018joint, balle18hyperprior}, and the VVC intra prediction mode (reference software VTM 18.0).
We first observe that our model achieves a much better PSNR-bpp curve than the previous theoretical bound of~\cite{yang2022sandwich} by 10\% in BD-rate, indicating that our results are closer to the true R-D function.
The improvement is also clear in terms of BD-rate, as shown in Table~\ref{table:rd_exp_bdrate}.
When compared to VTM 18.0, our bound is better by more than 30\% BD-rate on all three test sets.
This result is significant because it suggests that there is still a large room for improvement in lossy image compression.



\begin{table}[t]
\begin{adjustbox}{width=\linewidth}
\begin{tabular}{l|ccc}
\hline
                                       & \multicolumn{3}{c}{BD-rate (\%) w.r.t. VTM 18.0} \\
                                       & Kodak     & Tecknick    & CLIC   \\ \hline
\textit{Achievable bounds} \\
Ours                                   & \textbf{-31.4} & \textbf{-34.5} & \textbf{-31.9} \\
ResNet VAE~\cite{yang2022sandwich} (Yang and Mandt, 2021)\tablefootnote{The authors did not report results for the CLIC 2022 test set.} & -19.3     & -22.5       & -      \\
\hline
\textit{Learned image codecs} \\
QRes-VAE~\cite{duan2023qres} (Duan et al., 2023)          & -4.08     & -4.31       & -3.45  \\
TinyLIC~\cite{lu2022tinylic} (Lu et al., 2022)            & -3.10     & -5.06       & -5.57  \\
STF~\cite{zou2022stf} (Zou et al., 2022)                  & -2.09     & -6.22       & -2.48  \\
Joint AR \& H~\cite{minnen2018joint} (Minnen et al., 2018) & 11.5      & 9.37        & 14.1   \\
M \& S Hyperprior~\cite{balle18hyperprior} (Ballé et al., 2018) & 22.0      & 22.1        & 27.6   \\ \hline
\end{tabular}
\end{adjustbox}
\vspace{-0.4cm}
\caption{Comparison of methods in terms of BD-rate.}
\label{table:rd_exp_bdrate}
\vspace{-0.32cm}
\end{table}

\subsection{Additional Experiments}
\label{sec:rd_exp_additional}
We conduct an ablation study and show results in Table~\ref{table:rd_exp_model_config}.
We start from a small model and gradually increase its number of latent variables (rows 1-3) and its number of channels (rows 3-5).
The increase in model size improves the BD-rate, indicating that the model capacity is a key factor in estimating the R-D function.
Then, we train the smallest model and the largest model again but without our \ourfunction, which results in around 3\% BD-rate increases in both cases.
We thus conclude that our \ourfunction{} improves hierarchical VAEs with no additional computational complexity.

\begin{table}[t]
\footnotesize
\centering
\begin{tabular}{ccc|c|c}
\hline
\makecell{Smoothing \\ function}    & $C$ & $N$ & Params. & \makecell{Kodak BD-rate (\%) \\ w.r.t. VTM 18.0} \\ \hline
\checkmark & 64         & 5            & 34.2M                    & -8.98              \\
\checkmark & 64         & 10           & 46.1M                    & -16.5              \\
\checkmark & 64         & 15           & 55.8M                    & -19.7              \\
\checkmark & 96         & 15           & 111.8M                   & -28.8              \\
\checkmark & 128        & 15           & 186.7M                   & \textbf{-31.4}     \\ \hdashline
           & 64         & 5            & 34.2M                    & -5.88              \\
           & 128        & 15           & 186.7M                   & -28.7              \\
\hline
\end{tabular}
\vspace{-0.2cm}
\caption{Model configurations. $C$ is the number of channels used in Table~\ref{table:rd_method_config}, and $N$ is the total number of latent variables.}
\label{table:rd_exp_model_config}
\end{table}

\subsection{Discussion: Improving Image Codecs}
Given that existing image codecs are still far from optimum, how can we improve their R-D performance to approach the theoretical limit?
One possible solution is to develop Relative Entropy Coding~\cite{flamich2020rec, flamich2022a_star_rec} (REC) algorithms.
In the context of VAEs, REC algorithms aim at coding the latent variables with a rate close to $\mathbb{E} \left[ D_\text{KL}( q_{Z|X} \parallel p_Z ) \right]$.
Provided such an algorithm, our model could be directly turned into a practical compressor that achieves a rate-distortion performance close to $(U(D), D)$.
However, designing REC algorithms remains a challenging problem~\cite{theis2022algorithms_comm_sapmles}.
Another possible direction is to utilize the hierarchical architecture of ResNet VAEs which captures the coarse-to-fine nature of images, and preliminary works~\cite{duan2023qres, duan2023qarv} have shown promising results.




\section{Conclusion}
In this paper, we develop a new VAE architecture to obtain an upper bound for the rate-distortion function for lossy image compression.
Combined with novel training stabilization and variable-rate compression techniques, our model improves upon the existing upper bounding method by more than 10\% in BD-rate and reveals great potential for further improvements over current image compression methods (more than $30\%$ BD-rate reduction w.r.t. VVC intra prediction). 

\bibliographystyle{IEEEbib}
{\small \bibliography{references}}

\begin{thebibliography}{10}

\bibitem{pfaff2021vvc_intra}
J.~Pfaff, A.~Filippov, S.~Liu, X.~Zhao, J.~Chen, S.~De-Luxán-Hernández,
  T.~Wiegand, V.~Rufitskiy, A.~K. Ramasubramonian, and G.~Van der Auwera,
\newblock ``Intra prediction and mode coding in vvc,''
\newblock {\em IEEE Transactions on Circuits and Systems for Video Technology},
  vol. 31, no. 10, pp. 3834--3847, Oct. 2021.

\bibitem{cover2005information}
Thomas~M. Cover and Joy~A. Thomas,
\newblock {\em Elements of Information Theory},
\newblock John Wiley \& Sons, Inc., USA, 2006.

\bibitem{alemi2018fixing_elbo}
Alexander Alemi, Ben Poole, Ian Fischer, Joshua Dillon, Rif~A. Saurous, and
  Kevin Murphy,
\newblock ``Fixing a broken elbo,''
\newblock {\em Proceedings of the International Conference on Machine
  Learning}, vol. 80, pp. 159--168, July 2018.

\bibitem{yang2022sandwich}
Yibo Yang and Stephan Mandt,
\newblock ``Towards empirical sandwich bounds on the rate-distortion
  function,''
\newblock {\em International Conference on Learning Representations}, Apr.
  2022.

\bibitem{kingma14vae}
D.~Kingma and M.~Welling,
\newblock ``Auto-encoding variational bayes,''
\newblock {\em International Conference on Learning Representations}, Apr.
  2014.

\bibitem{admed1974dct}
N.~Ahmed, T.~Natarajan, and K.~Rao,
\newblock ``Discrete cosine transform,''
\newblock {\em EEE Transactions on Computers}, vol. C-23, no. 1, pp. 90--93,
  Jan. 1974.

\bibitem{wallace1992jpeg}
G.~Wallace,
\newblock ``The jpeg still picture compression standard,''
\newblock {\em IEEE Transactions on Consumer Electronics}, vol. 38, no. 1, pp.
  xviii--xxxiv, Feb. 1992.

\bibitem{zou2022stf}
Renjie Zou, Chunfeng Song, and Zhaoxiang Zhang,
\newblock ``The devil is in the details: Window-based attention for image
  compression,''
\newblock {\em Proceedings of the IEEE/CVF Conference on Computer Vision and
  Pattern Recognition}, pp. 17492--17501, June 2022.

\bibitem{lu2022tinylic}
Ming Lu and Zhan Ma,
\newblock ``High-efficiency lossy image coding through adaptive neighborhood
  information aggregation,''
\newblock {\em arXiv preprint arXiv:2204.11448}, Oct. 2022.

\bibitem{duan2023qres}
Zhihao Duan, Ming Lu, Zhan Ma, and Fengqing Zhu,
\newblock ``Lossy image compression with quantized hierarchical vaes,''
\newblock {\em Proceedings of the IEEE/CVF Winter Conference on Applications of
  Computer Vision}, pp. 198--207, Jan. 2023.

\bibitem{duan2023qarv}
Zhihao Duan, Ming Lu, Jack Ma, Yuning Huang, Zhan Ma, and Fengqing Zhu,
\newblock ``Qarv: Quantization-aware resnet vae for lossy image compression,''
\newblock {\em arXiv preprint arXiv:2302.08899}, Apr. 2023.

\bibitem{blahut1972rd}
Richard~E. Blahut,
\newblock ``Computation of channel capacity and rate-distortion functions,''
\newblock {\em IEEE Transactions on Information Theory}, pp. 460--473, July
  1972.

\bibitem{harrison2008rd}
Matthew~T. Harrison and Ioannis Kontoyiannis,
\newblock ``Estimation of the rate-distortion function,''
\newblock {\em IEEE Transactions on Information Theory}, pp. 3757--3762, Aug
  2008.

\bibitem{vahdat2020nvae}
Arash Vahdat and Jan Kautz,
\newblock ``Nvae: A deep hierarchical variational autoencoder,''
\newblock {\em Advances in Neural Information Processing Systems}, vol. 33, pp.
  19667--19679, Dec. 2020.

\bibitem{child2021vdvae}
Rewon Child,
\newblock ``Very deep vaes generalize autoregressive models and can outperform
  them on images,''
\newblock {\em International Conference on Learning Representations}, Apr.
  2021.

\bibitem{kingma2016iafvae}
Durk~P Kingma, Tim Salimans, Rafal Jozefowicz, Xi~Chen, Ilya Sutskever, and Max
  Welling,
\newblock ``Improved variational inference with inverse autoregressive flow,''
\newblock {\em Advances in Neural Information Processing Systems}, vol. 29,
  Dec. 2016.

\bibitem{sullivan1998rdo_video}
G.J. Sullivan and T.~Wiegand,
\newblock ``Rate-distortion optimization for video compression,''
\newblock {\em IEEE Signal Processing Magazine}, vol. 15, no. 6, pp. 74--90,
  Nov. 1998.

\bibitem{vaswani2017attention}
Ashish Vaswani, Noam Shazeer, Niki Parmar, Jakob Uszkoreit, Llion Jones,
  Aidan~N Gomez, \L~ukasz Kaiser, and Illia Polosukhin,
\newblock ``Attention is all you need,''
\newblock {\em Advances in Neural Information Processing Systems}, vol. 30,
  Dec. 2017.

\bibitem{ba2016layernorm}
Jimmy~Lei Ba, Jamie~Ryan Kiros, and Geoffrey~E Hinton,
\newblock ``Layer normalization,''
\newblock {\em arXiv preprint arXiv:1607.06450}, July 2016.

\bibitem{lin2014coco}
T.~Lin, M.~Maire, S.~Belongie, J.~Hays, P.~Perona, D.~Ramanan, P.~Doll{\'a}r,
  and C.~L. Zitnick,
\newblock ``Microsoft coco: Common objects in context,''
\newblock {\em Proceedings of the European Conference on Computer Vision}, pp.
  740--755, Sept. 2014.

\bibitem{Kingma15adam}
Diederik~P. Kingma and Jimmy Ba,
\newblock ``Adam: {A} method for stochastic optimization,''
\newblock {\em International Conference on Learning Representations}, May 2015.

\bibitem{kodak}
Eastman Kodak,
\newblock ``Kodak lossless true color image suite,''
  http://r0k.us/graphics/kodak/.

\bibitem{asuni2014tecnick}
Nicola Asuni and Andrea Giachetti,
\newblock ``{TESTIMAGES: a Large-scale Archive for Testing Visual Devices and
  Basic Image Processing Algorithms},''
\newblock {\em Smart Tools and Apps for Graphics - Eurographics Italian Chapter
  Conference}, Sept. 2014.

\bibitem{bjontegaard2001bdrate}
Gisle Bjontegaard,
\newblock ``Calculation of average psnr differences between rd-curves,''
\newblock {\em Video Coding Experts Group - M33}, Apr. 2001.

\bibitem{minnen2018joint}
D.~Minnen, J.~Ball\'{e}, and G.~Toderici,
\newblock ``Joint autoregressive and hierarchical priors for learned image
  compression,''
\newblock {\em Advances in Neural Information Processing Systems}, vol. 31, pp.
  10794--10803, Dec. 2018.

\bibitem{balle18hyperprior}
J.~Ballé, D.~Minnen, S.~Singh, S.~Hwang, and N.~Johnston,
\newblock ``Variational image compression with a scale hyperprior,''
\newblock {\em International Conference on Learning Representations}, Apr.
  2018.

\bibitem{flamich2020rec}
Gergely Flamich, Marton Havasi, and Jos\'{e}~Miguel Hern\'{a}ndez-Lobato,
\newblock ``Compressing images by encoding their latent representations with
  relative entropy coding,''
\newblock {\em Advances in Neural Information Processing Systems}, vol. 33, pp.
  16131--16141, Dec. 2020.

\bibitem{flamich2022a_star_rec}
Gergely Flamich, Stratis Markou, and Jos{\'e}~Miguel Hern{\'a}ndez-Lobato,
\newblock ``Fast relative entropy coding with a* coding,''
\newblock {\em Proceedings of the International Conference on Machine
  Learning}, vol. 162, pp. 6548--6577, July 2022.

\bibitem{theis2022algorithms_comm_sapmles}
Lucas Theis and Noureldin~Y Ahmed,
\newblock ``Algorithms for the communication of samples,''
\newblock {\em Proceedings of the International Conference on Machine
  Learning}, vol. 162, pp. 21308--21328, June 2022.

\end{thebibliography}
\end{document}